# EEG-based Cross-Subject Driver Drowsiness Recognition with an Interpretable Convolutional Neural Network

Jian Cui, Zirui Lan, Olga Sourina, Wolfgang Müller-Wittig

*Abstract*—In the context of electroencephalogram (EEG)-based driver drowsiness recognition, it is still challenging to design a calibration-free system, since EEG signals vary significantly among different subjects and recording sessions. Many efforts have been made to use deep learning methods for mental state recognition from EEG signals. However, existing work mostly treats deep learning models as black-box classifiers, while what have been learned by the models and to which extent they are affected by the noise in EEG data are still underexplored. In this paper, we develop a novel convolutional neural network combined with an interpretation technique that allows sample-wise analysis of important features for classification. The network has a compact structure and takes advantage of separable convolutions to process the EEG signals in a spatial-temporal sequence. Results show that the model achieves an average accuracy of 78.35% on 11 subjects for leave-one-out cross-subject drowsiness recognition, which is higher than the conventional baseline methods of 53.40%-72.68% and state-of-the-art deep learning methods of 71.75%-75.19%. Interpretation results indicate the model has learned to recognize biologically meaningful features from EEG signals, e.g., Alpha spindles, as strong indicators of drowsiness across different subjects. In addition, we also explore reasons behind some wrongly classified samples with the interpretation technique and discuss potential ways to improve the recognition accuracy. Our work illustrates a promising direction on using interpretable deep learning models to discover meaningful patterns related to different mental states from complex EEG signals.

## I. INTRODUCTION

Causing decrease in attention, vigilance and cognitive performance, driver's drowsiness is a leading factor of car accidents. Development of a drowsiness monitoring system to continuously watch the vigilance state of the driver and send alarm before the driver falls asleep is of high priority for driving safety and prevention of transportation accidents. Many efforts have been made to investigate monitoring of driver drowsiness using electroencephalogram (EEG), which is believed to be the most practical non-invasive modality for capturing brain dynamics due to its high temporal resolution and low cost. However, building a calibration-free drowsiness recognition system is still a challenging task. The difficulty lies in capturing common drowsiness-related patterns from a diversity of EEG signals with a low signal-to-noise rate. The variability of EEG signals from different subjects is attributed to many factors, such as electrode displacements, skin-electrode impedance, different head shapes and sizes, different brain activity patterns, and disturbance by task-irrelevant brain activities. Conventional methods relying on hand-crafted features are often very specific to some EEG characteristics of interest, which potentially excludes other relevant information that could be essential to drowsiness recognition. In comparison, deep learning allows end-to-end learning without the need for a priori feature crafting. Such models can directly learn essential characteristics from raw high-dimensional data by converting it into a cascade of representations while optimizing the parameters through back propagation. However, existing work in the area of EEG signal processing mostly treats deep learning models as black-box classifiers, while what have been learned by the models and to which extent they are affected by the noise from EEG data are still underexplored. Without knowing these facts, the work of developing the model towards higher accuracy becomes a trial-and-error process.

In this paper, we propose a novel Convolutional Neural Network (CNN) named "InterpretableCNN" for driver drowsiness recognition and discovering common drowsiness-related patterns of EEG signals across different subjects. The network has a compact structure and it uses separable convolutions to process the EEG signals in a spatial-temporal sequence. In order to allow the model to "explain" its decisions, we have designed a novel interpretation technique for the model that can reveal local regions of the input signals that are important for prediction.

In the following part of this paper, existing works are reviewed in Section II. The methods are proposed in Section III. The performance of the proposed method is evaluated in Section IV, which is followed by discussion and future works in Section V. Conclusions are made in Section VI.

## II. RELATED WORKS

### A. Existing driver monitoring systems

Driver monitoring system is a safety system installed on vehicles for monitoring the alertness of drivers and making alarm when the driver gets drowsy or falls asleep. The first driver monitoring system was introduced by Toyota in 2006 for its Lexus latest models [1]. The system uses a CCD camera placed on the steering column to track face of the driver via infrared LED detectors. It alarms the driver by flashing lights or warning sounds when the driver is drowsy or not paying attention to the road ahead. Ford introduced the Focus and Mondeo models that were equipped with driver alert systems [2], which use a forward-looking camera to monitor the vehicle's position in relation to the lane and infer the fatigue state of the driver through his/her lane-tracking performance. Ford developed a biometric car seat prototype to monitor the stress and inattention of drivers with multiple physiological signals [3]. Specifically, the prototyped car seat features a seatbelt with integrated piezoelectric film to monitor respiration. Infrared sensors are placed on the steering wheel spokes to monitor the driver's facial temperature. Conductive sensors built into the rim of the steering wheel can monitor the heart rate and skin conductivity of the driver when those contact points are grabbed. Nissan launched the Maxima model with a system named Driver Attention Alert, which can track the driver's steering pattern using the steering angle sensor and predict the act of drowsy driving [4].

In summary, camera-based, driver behavior-based and physiological signal-based methods are primarily used by car manufacturers for drowsiness detection. In comparison to existing methods, EEG has the advantage of directly monitoring brain activities with a high temporal resolution, so that it allows drowsiness to be detected in an early stage before it is reflected from the driver behaviors [5].

### B. EEG-based driver drowsiness recognition

EEG measures voltage difference on the scalp, which is resulted from ion currents caused by synaptic activities of thousands of pyramidal neurons happening on the surface layer of the brain underneath [6]. Existing EEG systems have 1 to 256 electrodes, and the placement of electrodes follows a formal standard called 10–20 system or International 10–20 system. EEG channels follow an alphanumeric naming convention. The channel name starts with one- or two-letter acronym and ends with a number or letter "z", e.g., AF3, Cz, and T4. Letter Fp, F, T, P, O, and C stand for pre-frontal, frontal, temporal, parietal, occipital and central, respectively. The ending digit denotes the placement of the electrode on the coronal line, while "z", standing for zero, denotes the center of the coronal line.

Existing studies have revealed the relationship between drowsiness and the oscillation patterns of EEG signals. For example, Akerstedt et al. [7] found a significant increase in power from the Theta (4-7.9 Hz) and Alpha (8-11.9 Hz) frequency bands for subjects during night driving in comparison to those in day driving. Corsi-Cabrera et al. [8] found an increase in power from frequency bands of fast upper Alpha (9.77-12.45 Hz) and Beta (12.7-17.85 Hz) from subjects who experienced sleep deprivation. In another experiment conducted in a driver simulator [9], the subjects were found to have higher power in the EEG frequency bands during the early sleep stage. It was summarized by Klimesch [10] that drowsiness can in general cause an increase in power of the Theta and Alpha frequency bands. They further concluded that the increase in lower Alpha power occurs only when subjects are struggling not to sleep, while the Alpha power will decrease when subjects fall asleep.

With the emergence of deep learning, the performance and accuracy of many artificial intelligence and classification tasks have been vastly boosted. In comparison to traditional EEG processing methods based on feature crafting, deep learning can directly learn from raw data and transform it into a cascade of representations with increasing abstraction. The important characteristics of raw EEG signals can therefore be retained when the processes of feature extraction and classifier training are combined under the same learning framework. For the task of driver drowsiness recognition, Nissimagoudar et al. [11] proposed a simple convolutional network consisting of two convolutional layers, a max-pooling layer, a flatten and a fully connected layer to classify single-channel EEG data for Advance Driver Assistance Systems (ADAS) of automotive. Ding et al. [12] implemented a deep CNN model on a mobile device to detect drowsiness from single-channel EEG signals. The model employs cascaded CNN and attention mechanism layer in the structure. Results show that their model outperforms other benchmark deep learning models, as well as conventional machine learning methods, such as Support Vector Machine (SVM) and Linear Discriminant Analysis (LDA). In order to process multi-channel EEG signals, Gao et al. [13] proposed a model called EEG-based Spatial–Temporal CNN (ESTCNN) for driver fatigue detection. The model contains three core blocks, and each block has a convolution layer, a rectified linear activation layer, a flatten and a batch normalization layer. Zeng et al. [14] developed two CNN models called EEG-Conv and EEG-Conv-R, respectively. The first one is based on the traditional CNN and second one combines CNN with deep residual learning. They found both models outperform the Long Short-Term Memory (LSTM)- and SVM-based classifiers, while EEG-Conv-R converges more quickly.

Despite the promising results reported in existing work on using deep learning for EEG-based driver drowsiness recognition, there is still little insight on what characteristics of the EEG data have been learned by the models from the data. In fact, interpretation on what have been learned by the models is an important procedure for validation of the results, since it not only allows us to find out whether the classification is driven by relevant features in the data, but also potentially discover interesting neurophysiological patterns associated with different mental states. In our previous work [15, 16], initial attempts were made to design interpretable deep models for driver drowsiness recognition from single-channel EEG signals. In this paper, we propose a compact CNN model to deal with multi-channel EEG signals, which is combined with a novel visualization technique that enables the model to "explain" its decisions.

### III. MATERIALS AND METHODS

### A. Data preparation

A public EEG dataset [17] was used in the study. The dataset was collected from 27 subjects (aged from 22 to 28) in a sustained-driving task in a virtual reality simulator. In the process, lane-departure events were randomly introduced which drifted the car away from the central lane. The participants were required to respond immediately to the events by driving the car back to central lane. The drowsiness level can be reflected by how fast the subjects respond to the events.

The EEG signals were sampled at 500 Hz with 30 electrodes and processed with 1-50 Hz bandpass filters and artifact rejection. The pre-processed version of the dataset available from [18] was used in this study. We further down-sampled the original data to 128 Hz and extracted the EEG samples of 3-second length prior to the car deviation events for each trail. Each sample has a dimension of 30 (channels) × 384 (sample points). We followed methods described in [19] to select and label the EEG samples. Specifically, the local reaction time (RT), which is the time taken by the subject to respond to the car drift event, and the global-RT, which is the average of RTs within a 90-second window before the car drift event, were calculated for each sample. The baseline "alert-RT" for each session was defined as the 5th percentile of the local RTs. Samples with both local-RT and global-RT shorter than 1.5 times alert-RT were labeled as alert state, while samples with both local-RT and global-RT longer than 2.5 times alert-RT were labeled as drowsy state. The complete table of the number of labeled samples for each session can be found from [15].

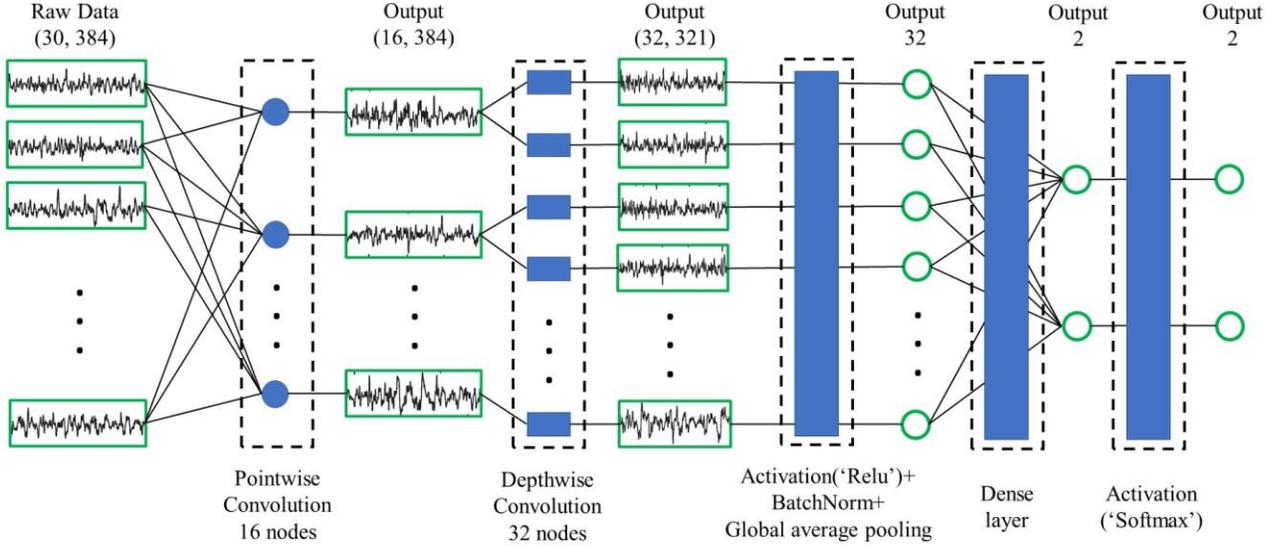

Fig. 1. The architecture of the proposed model. The shapes with green outlines are the input signals or outputs from the intermediate layers. The blue shapes inside the dashed borders represent kernels or layers of the network. For the tuples above the figure, the first entry indicates the channel number and the second entry indicates the length of the input signal or the intermediate data.

We further discarded sessions with less than 50 samples of either class. If there were multiple sessions of the same subject, the session with the most balanced class distribution for the subject was used. We performed these two additional selection procedures in order to loosely balance the samples for each class and each subject, so that the classifiers will not have to predict data from a specific subject. In this way, we finally got an unbalanced dataset of 2952 samples from 11 different subjects, and the number of samples from each subject is shown in Table I.

TABLE I. NUMBER OF EXTRACTED SAMPLES FROM EACH ELIGIBLE SUBJECT

| Subject ID | Sample Number | |
|---|---|---|
| | Alert | Drowsiness |
| 1 | 94 | 96 |
| 2 | 363 | 66 |
| 3 | 75 | 180 |
| 4 | 118 | 74 |
| 5 | 161 | 112 |
| 6 | 83 | 116 |
| 7 | 51 | 103 |
| 8 | 238 | 132 |
| 9 | 243 | 157 |
| 10 | 192 | 54 |
| 11 | 113 | 131 |
| Total | 1731 | 1221 |

Additionally, we have also built a balanced dataset from the unbalanced dataset by choosing the most representative samples from the majority class for each subject with shortest (for alert class) or longest (for fatigue class) local-RTs. The balanced dataset is ideal for training the classifiers, while the unbalanced dataset is closer to the real-life situation. Both of the balanced dataset and the unbalanced dataset have been uploaded online [20, 21].

### B. Network design

#### 1) The core idea

The EEG signals can be viewed as a mixture of cortical source signals generated from different areas of the brain. However, the recorded data are inevitably contaminated by artifacts caused by different electrical activities, e.g., cardiac, eye movement and muscle tension, as well as noise generated from the equipment. Considering learning directly from the noisy and redundant EEG data usually leads to unsatisfactory recognition results, spatial filtering techniques [22] were proposed to improve the data quality by extracting a set of new signals from the raw multi-channel recordings of EEG with minimal contamination and redundancy. In this section, we start from the spatial filtering techniques and then introduce the network design inspired by the processing flows.

Suppose the EEG signals recorded from $m$ electrodes are $\{x_i\}_{i=1,2\ldots m}$. $N_1$ new signals $\{s_j\}_{j=1,2\ldots N_1}$ can be obtained from linear combination of the original $m$ signals.

$$s_j = \sum_{i=1}^{m} w_{i,j} x_i + b_j \qquad (1)$$

In (1), the weights $\{w_{i,j}\}$ are a set of spatial filters, which can be calculated based on various independent evaluation criteria such as distance measure, information measure, dependency measure, and consistency measure [22]. For example, Independent Component Analysis (ICA) [23] is a well-known method that finds the weights by solving an equation based on the statistical independency hypothesis of the source signals. Other methods include Common Spatial Pattern (CSP) [24], Minimal Energy Combination (of noise), Maximum Contrast Combination (MCC) [25], Canonical Correlation Analysis (CCA) [26], and so forth. Since the obtained new set of signals $\{s_j\}$ are expected to contain minimal noise and redundancy, a set of features can be thus extracted from each new signal $s_j$ to be used for classification.

$$[feature_{j,1}\ feature_{j,2}\ ...,feature_{j,k}] = f_j(s_j) \quad (2)$$

For driver drowsiness recognition, Lin *et al.* [27] proposed to use the feature extraction pipeline, where the raw EEG data were demixed with the ICA method and band power features were obtained from each of the demixed signals. In our model, we have incorporated the processing pipeline in the CNN structure allowing the parameters to be optimized towards a higher recognition accuracy.

### 2) The context of separable convolution

The proposed processing sequence of EEG signal is similar to the concept of depthwise separable convolution, which is an operation proposed as early as 2014 [28] and adopted in many network structures, e.g., MobileNets [29, 30], with the purpose of reducing parameters and encouraging convergence of the network.

A depthwise separable convolution, or called "separable convolution", consists of a depthwise convolution, which is performed on each channel independently, and a pointwise convolution (or called "1x1 convolution"), which projects the channels onto a new space. In the context of image classification, the depthwise separable convolution is different from a standard 2D convolution in the aspect that it independently learns the spatial correlations (consisting of width and height of an image) and the cross-channel correlations with the depthwise and pointwise convolutions, respectively, instead of simultaneously learning the parameters in a 3D space.

For our case, we use the pointwise convolution to implement the first processing step (1) of demixing the signals and the depthwise convolution to implement the second processing step (2) of learning temporal features from each demixed signal independently. Our model is different from the typical depthwise separable convolution [29, 30] in the aspects of the operation order, while it is almost identical to the "extreme" version of the inception module used in a deep learning model named Xception, which was proposed by Chollet [31].

### 3) Network structure

The network consists of seven layers and its structure is shown in Fig. 1. The pointwise convolution and depthwise convolution are implemented in the first two layers, which are followed by a ReLU activation layer, a batch normalization layer, a global average pooling layer, a dense layer and a Softmax activation layer.

In the first layer, we use $N_1$ pointwise convolutional nodes to generate $N_1$ new channels of signals with the same length. For a given input sample $X_{(m,n)}$, which contains $m$ (=30) channels of 1-dimensional signal with length of $n$ (=384). The $i$-th ($1 \leq i \leq N_1$) pointwise convolutional node has $m$ weights $\{w_{i,p}^{(1)}\}_{p=1,2,...,m}$ and 1 bias $b_i^{(1)}$ parameter. The outputs from the first layer is

$$h_{i,j}^{(1)} = \sum_{p=1}^{m} w_{i,p}^{(1)} x_{p,j} + b_i^{(1)}, \quad (3)$$

where $i$ = 1, 2, 3, …, $N_1$ and $x_{p,j}$ is the $j$-th sampling point of the $p$-th channel of the input EEG sample. The superscripts of the outputs and network parameters indicate which network layer they belong to. We set $N_1$=16, which is around half length of the input channels, in order to reduce redundancy and encourage convergence of the network.

In the second layer, depthwise convolutions are used to extract features from the $N_1$ obtained signals. Specifically, each new signal $h_i^{(1)}$ is convoluted with two nodes in the second layer. Since there are $N_1$ channels of new signals output from the first layer, we have in total $2N_1$ depthwise convolutional nodes. Suppose the length of the kernel is $l$, the $i$-th ($1 \leq i \leq 2N_1$) node has $l$ weights $\{w_{i,r}^{(2)}\}_{1 \leq r \leq l}$ and 1 bias $b_i^{(2)}$ parameter. The output of the layer is

$$h_{i,j}^{(2)} = \begin{cases} \sum_{r=1}^{l} h_{\frac{i+1}{2},j+r-1}^{(1)} w_{i,r}^{(2)} + b_i^{(2)}, when\ i\ is\ odd. \\ \sum_{r=1}^{l} h_{\frac{i}{2},j+r-1}^{(1)} w_{i,r}^{(2)} + b_i^{(2)}, when\ i\ is\ even. \end{cases} \quad (4)$$

In (4), the length of a kernel is set as $l = 64$, which is half of the sampling rate (128 Hz). The size of the output $h_{i,j}^{(2)}$ is ($2N_1$, $n-l+1$), which is (32, 321). The 3rd and 4th layers are activation and batch normalization layers.

$$h_{i,j}^{(3)} = ReLU(h_{i,j}^{(2)}) \quad (5)$$

$$h_{i,j}^{(4)} = BatchNorm(h_{i,j}^{(3)}) \quad (6)$$

Global Average Pooling (GAP) [32] is used in the 5th layer. In comparison to the widely used fully connected layer, the GAP layer dramatically reduces parameters and can thus effectively prevent over-fitting.

$$h_i^{(5)} = \left(\sum_{j=1}^{n-l+1} h_{i,j}^{(4)}\right)/(n - l + 1) \quad (7)$$

The model ends with a dense layer and a Softmax activation layer.

$$h_c^{(6)} = \sum_{i=1}^{2N_1} w_{i,c}^{(6)} h_i^{(5)} + b_c^{(6)} \quad (8)$$

$$h_c^{(7)} = Softmax(h_c^{(6)}) \quad (9)$$

In (8) and (9), $c$ = 0 or 1, which represents the alert or drowsy state, respectively. We name the proposed model "InterpretableCNN" considering the classification results of the model can be interpreted with the technique proposed in the next section, which is an advantage over other state-of-the-art deep learning models.

### C. Interpretation technique

Deriving insights on what characteristics of EEG signals are learned by the deep learning networks has become an important procedure of model validation, since it not only allows us to find out whether the classification is driven by relevant features in the data, but also potentially discover interesting neurophysiological patterns associated with different mental states. In existing literature, attempts have been made to interpret deep learning models for EEG signal classification by visualizing kernel weights, summarizing averaged output from hidden unit activations, calculating single-trial feature relevance [33], reconstruction of the sample that leads to maximal activations [34], and so forth. Although these techniques allow understanding of what global patterns have been learned from the massive data, they cannot explain what specific characteristics of each EEG sample are found relevant to different mental states and to which extent the models are affected by noise from the data.

By comparison, the Class Activation Map (CAM) [35] method is a powerful interpretation technique that can localize the discriminative regions of each input sample for a CNN model trained to solve a classification task. Specifically, for each input sample a heatmap is generated from the activations after the last convolutional layer. The map is then interpolated to size of the input sample and it reveals to which extent the local regions of the input sample contribute to the classification. However, since the CAM method was originally designed for deep CNN networks with only standard convolutional layers (e.g., Alexnet [36]) for classification of image data, it cannot be directly used for the proposed model as well as other CNN models involving convolutions in the spatial dimensional [13] or having structures more than basic convolutions blocks [33]. Therefore, we begin with the CAM method and design a novel interpretation technique for the proposed model.

Suppose a given input EEG sample $X_{(m,n)}$ is classified with label $c$, where $c$ is either 0 or 1 representing the alert or drowsy state, respectively. The objective is to find the heatmap $S_{(m,n)}^c$ for $X_{(m,n)}$ that can reveal important regions for the prediction by the network. Suppose an input signal $X_{(m,n)}$ generates activation $h_c^{(6)}$ in the 6th layer of the network. From (7) and (8), we have

$$h_c^{(6)} = \sum_{i=1}^{2N_1} w_{i,c}^{(6)} h_i^{(5)} = \sum_{i=1}^{2N_1} \sum_{j=1}^{n-l+1} w_{i,c}^{(6)} h_{i,j}^{(4)}$$
$$= \sum_{i=1}^{2N_1} \sum_{j=1}^{n-l+1} M_{i,j}^c , \qquad (10)$$

where

$$M_{i,j}^c = w_{i,c}^{(6)} h_{i,j}^{(4)} \qquad (11)$$

$M_{i,j}^c$ is the activation map of class c for the sample $X_{(m,n)}$, as defined in [35]. In (10), we neglect the constants $b_c^{(6)}$ and $(n - l + 1)$ for simplicity. As it can be seen in (10), $M_{i,j}^c$ can be viewed as the distribution of the final activation $h_c^{(6)}$ for class $c$ in a map of size $2N_1 \times (n - l + 1)$. The original CAM method finds the heatmap by upsampling $M_{i,j}^c$ until it has the same size as the input sample. However, the method cannot be directly used for our model since the channels of the input signal are mixed by pointwise convolutions in the first layer, which makes the first (channel) dimension of $M_{i,j}^c$ misaligned with the first

(channel) dimension of the input signal. Inspired by the CNN-Fixation method [37], we consider an alternative way of tracing only a small portion of the positions in the activation map $M_{i,j}^c$ that contribute most to the class activation $h_c^{(6)}$ rather than the whole activation map, back to the their major corresponding areas in the input signal. Specifically, we rank the values of $M_{i,j}^c$ in a descending order. Suppose the locations of the first $N$ elements in $M_{i,j}^c$ are $(i_0, j_0), (i_1, j_1), ..., (i_N, j_N)$, where $1 \le i_k \le 2N_1$ ($2N_1 = 32$) and $1 \le j_k \le n - l + 1$ ($n - l + 1 = 321$). The objective is to trace each of these discriminative locations for class $c$ in $M_{i,j}^c$ throughout the network to the center of areas in the input sample that contribute most to these high activations. Suppose we can find $N$ corresponding discriminative locations in the input samples and they are $(p_0, q_0), (p_1, q_1), ..., (p_N, q_N)$, where $1 \le p_k \le m$ ($m = 30$) and $1 \le q_k \le n$ ($n = 384$). The final heatmap for sample $X_{(m,n)}$ can be obtained by combining all the class discriminative points in the input sample with the Gaussian function.

$$S_{p,q}^c = \frac{1}{\sigma \sqrt{2\pi}} \sum_{k \; for \; p_k=p} e^{\left(-\frac{1}{2} \frac{(q-q_k)^2}{\sigma^2}\right)} \qquad (12)$$

In (12), $\sigma$ is a constant that decides radius of the influential area of each discriminative point in the input signal. $S_{p,q}^c$ is further normalized in the range (-1, 1) for visualization.

Finally, we consider how to trace the set of discriminative locations $(i_0, j_0), (i_1, j_1), ..., (i_N, j_N)$ in $M_{i,j}^c$ back to the input sample through the four layers (layer 1-4) of the proposed network. It is easy to notice that the discriminative locations are unchanged after the 3rd and 4th layers of the network, since the activation and batch normalization layers only perform element-wise operations that will not affect the topology of the data. Therefore, the only task left for us is to consider how to trace the discriminative locations through the depthwise and pointwise convolutional layers, which was not discussed in the original CNN-Fixation method [37]. Suppose the input sample $X_{(m,n)}$ generates activation $h_{i_k,j_k}^{(2)}$ after the 2nd layer of the network at the discriminative location $(i_k, j_k)$. From (3) and (4), we have

$$h_{i_k,j_k}^{(2)} = \sum_{r=1}^{l} h_{\frac{i_k+1}{2}, j_k+r-1}^{(1)} w_{i_k,r}^{(2)}$$
$$= \sum_{r=1}^{l} w_{i_k,r}^{(2)} \sum_{p=1}^{m} w_{\frac{i_k+1}{2}, p}^{(1)} x_{p, j_k+r-1}$$
$$= \sum_{p=1}^{m} w_{\frac{i_k+1}{2}, p}^{(1)} \sum_{r=1}^{l} w_{i_k,r}^{(2)} x_{p, j_k+r-1} , \quad (13a)$$

when $i_k$ is odd, and similarly

$$h_{i_k,j_k}^{(2)} = \sum_{p=1}^{m} w_{\frac{i_k}{2}, p}^{(1)} \sum_{r=1}^{l} w_{i_k,r}^{(2)} x_{p, j_k+r-1} , (13b)$$

when $i_k$ is even. In (13a) and (13b), we ignore $b_i^{(1)}$ for simplicity of expression. From (13a, b), we can observe that $h_{i_k,j_k}^{(2)}$ is generated from an episode of the input signals at the local area from the time point $j_k$ to $j_k + l - 1$. Actually, it is the weighted sum of the convoluted signals $\sum_{r=1}^{l} w_{i_k,r}^{(2)} x_{p, j_k+r-1}$ of all the $m$ channel, and the weight assigned to the channel $p$ is $w^{(1)}_{(i_k+1)/2,p}$ or $w^{(1)}_{i_k/2,p}$. Therefore, the discriminative location $(i_k, j_k)$ in $M^c_{i,j}$ can be traced back to the center $(p_k, q_k)$ of the strongest contributing episode in the input signal, where

$$p_k = argmax_p \left( w^{(1)}_{\frac{i_k+1}{2},p} \sum_{r=1}^{l} w^{(2)}_{i_k,r} x_{p,j_k+r-1} \right) \quad (14a)$$

when $i_k$ is odd,

$$p_k = argmax_p \left( w^{(1)}_{\frac{i_k}{2},p} \sum_{r=1}^{l} w^{(2)}_{i_k,r} x_{p,j_k+r-1} \right) \quad (14b)$$

when $i_k$ is even, and

$$q_k = j_k + (l-1)/2 \quad (15)$$

We set $\sigma = l/2 = 32$ for (12), so that the discriminative location in the input signal will highlight the whole episode of strongest contributing signal. We trace the top 100 ($N$=100) discriminative locations in class activation map $M^c_{i,j}$, which accounts for around 1% of all entries of $M^c_{i,j}$.

### D. Methods for comparison

In this part, we compare the performance of the proposed model with both state-of-art deep learning and conventional baseline methods. In addition, we also compare the model with its variations to understand how each component of the model influences its overall performance

#### 1) Deep learning methods

The first deep learning model we use for comparison is the benchmark CNN model for EEG signal classification—EEGNet proposed by Lawhern et al. [33]. Inspired by the filter bank common spatial patterns (FBCSP) algorithm [38], EEGNet uses a 2D convolutional layer to filter the raw signals, which is followed by a spatial depthwise convolutional layer and a temporal depthwise convolutional layer to extract features. The model was tested on several Brain Computer Interface (BCI) datasets and achieved higher accuracies than conventional methods. The model has also been tested for cross-subject driver drowsiness recognition in preliminary studies conducted by Liu et al. [39, 40]. In this study, two configurations of EEGNet–EEGNet-4,2 and EEGNet-8,2 were used as baseline deep learning methods for comparison.

The second deep learning model we use for comparison is the Sinc-ShallowNet model proposed by Davide et al. [41]. The model has a sinc-convolutional layer and a depthwise convolutional layer to process the raw EEG data in a temporal-spatial sequence. The sinc-convolutional layer, which was initially proposed by Ravanelli and Bengio [42], is different from a standard convolutional layer in the aspect that it uses sinc functions parametrized with only two cutoff frequencies forming band-pass filters as kernels of the layer. In addition, we also implement a variation of the model named "Conv-ShallowNet", which replaces the sinc-convolutional layer with a standard convolutional layer for comparison.

#### 2) Conventional baseline methods

Conventional methods for EEG signal classification mainly involve the stages of feature extraction and feature classification. In order to have a comprehensive understanding on the performance of different conventional methods on the dataset, we implement five baseline methods for feature extraction and test them on eight different classifiers for comparison.

EEG band power features have been regarded as golden standard for EEG signal classification. For driver drowsiness recognition, many works [8, 9] have found a strong relationship between drowsiness and band power features of EEG signals. Therefore, the first three baseline methods use different forms of band power features, which are relative band power features, log of band power features [43], and the ratio of band power features [44]. The fourth baseline method uses the wavelet entropy features [45], while the fifth baseline method uses a combination of four entropies features [46], which are sample entropy, fuzzy entropy, approximate entropy and spectral entropy. Details of these methods are illustrated below.

**RelativePower**: we use relative band power features as the first baseline method for comparison. Specifically, power features from four frequencies bands of Delta (1–4 Hz), Theta (4–8 Hz), Alpha (8–12 Hz) and Beta (12–30Hz) are extracted from each EEG channel. Considering the absolute values of the band powers vary significantly across different samples, relative power of the four frequency bands from each EEG channel are calculated.

**LogPower**: the second baseline method is slightly different from the first one—natural log of the band power instead of relative power is calculated. The method was proposed by Pal et al. [43] based on the observation of a strong linear correlation between log power features of EEG and subject's driving performance.

**PowerRatio**: Jap et al. [44] found four band power ratios (i) $(\theta + \alpha)/\beta$, (ii) $\alpha/\beta$, (iii) $(\theta + \alpha)/(\alpha + \beta)$ and (iv) $\theta/\beta$ were good indicators of driver drowsiness. Therefore, for the third baseline method, we calculate the four band power ratio features and use them as representations of the EEG sample signals.

**WaveletEntropy**: we implement the method proposed by Wang et al. [45], where wavelet entropy features from EEG signals are used to recognize driver drowsiness. Specifically, the Mexican Hat Wavelet is used in our implementation, and wavelet coefficients on the wavelet scales of 0.5, 1, 2, 4, 8, 16, 32 (corresponding to frequencies of 64 Hz, 32 Hz, 16 Hz, 8 Hz, 4 Hz, 2 Hz, 1 Hz) are extracted from each EEG channel. The wavelet entropy feature for each EEG channel is calculated by applying the Shannon function on the normalized wavelet coefficients.

**FourEntropies**: Hu et al. [46] proposed to use four types of entropies, which are sample entropy, fuzzy entropy, approximate entropy and spectral entropy for driver fatigue recognition. Following the descriptions in the paper, we calculated the approximate entropy and spectral entropy using the methods proposed by Song et al. [47] and the fuzzy entropy by the method proposed by Xiang et al. [48], and set the parameters $m$ and $r$ involved in the calculation as $m = 2$ and $r = 0.2 * SD$. We set the width of the exponential function $n$ as $n$=2 for extracting fuzzy entropy features. Finally, we normalize each feature dimension for each subject, as indicated in the original paper.

**Classifiers**: Different classifiers have been implemented, which include Decision Tree (DT), Random Forest (RF), k-

nearest neighbors (KNeighbors), Gaussian Naive Bayes (GNB), Logistic Regression (LR), LDA, Quadratic Discriminant Analysis (QDA), and SVM.

### 3) Variations of the model for comparison

The proposed model is compared with its variations in order to understand how each part of the model influences its overall performance.

Firstly, in order to evaluate the benefits of using the separable convolution over the standard 1-dimensional convolution, we replace the pointwise and the depthwise convolutional layers of the network with a standard convolutional layer containing 32 kernels with length of 64. We name this model as "1DConv".

In the second and the third variations, we remove the depthwise convolutional layer and the pointwise convolutional layer, respectively, in order to understand how these two layers contribute to the final classification. We name these model as "NoDepthwise" and "NoPointwise", repectively.

Finally, we consider a variation of the model, where the batch normalization layer is removed. We name the model as "NoBatchNorm".

### E. Implementation details

The comparison was conducted on an Alienware Desktop with 64-bit Windows 10 operation system powered by Intel(R) Core(TM) i7-6700 CPU and an NVIDIA GeForce GTX 1080 graphics card. The codes were implemented and tested on the platform of Python 3.6.6. The proposed model, the Sinc-ShallowNet model and their variations were implemented with the Pytorch Library. The EEGNet models were downloaded from [49] and run with the Keras API of TensorFlow.

Our initial tests showed that all these deep learning models have a fast converge followed by a significant drop on the accuracies for this dataset when using the default batch normalization layer provided by both the Pytorch and TensorFlow libraries. In order to solve the problem, we conducted minor modifications on all the comparison models by disabling estimation of the computed mean and variance (we set *track_running_stats=False* and *momentum=0* for the batch normalization layers implemented by the Pytorch and Tensflow libraries, respectively). For training of the neural network models, although different optimizers have been used in various deep architectures, such as Sobolev gradient based optimizers [50, 51], we used Adam [52] due to its low computational cost and efficiency in our technique. We set batch size as 50 and used default parameters ($\eta = 0.001$, $\beta_1 = 0.9$, $\beta_2 = 0.999$) for optimization with the Adam method. The source codes of the proposed model and the interpretation technique are accessible from [53].

As for the conventional methods, band power features were extracted using the Welch method from the SciPy library [54]. The classifiers were implemented with the sklearn library [55] and the default parameters were used.

## IV. EVALUATION ON THE PROPOSED METHOD

### A. Model comparison results

In this section, we conduct leave-one-subject-out cross validation to compare the classifiers. Specifically, the EEG data from one subject are used for testing, while data from all the other subjects are used for training the classifiers. The process is iterated until every subject serves once as the test subject.

We consider the ideal case in Section 1), where the balanced dataset is used for a standard evaluation on the performance of different classifiers. In Section 2), we compare selected classifiers on the unbalanced dataset, which is closer to the real-life case.

### 1) Mean accuracy comparison on the balanced dataset

In this part, we compare the mean accuracies of different classifiers on the balanced dataset. We trained each deep learning model from 1 to 50 epochs. Considering neural networks are stochastic, we randomized the network parameters for each iteration and repeated the process for 10 times. In this way, 10 (times) x 11 (subjects) = 110 folds were created for each epoch.

The mean classification accuracies of the proposed model InterpretableCNN and the benchmark deep learning models against training epochs from 1 to 50 are shown in Fig. 2(a). As it can be seen in the figure, InterpretableCNN has an overall better performance than the other four models. It reaches the peak accuracy of 78.35% in 11 epochs, after which the accuracy drops a little but still stabilizes at above 76% in the rest of the first 50 epochs. EEGNet-4,2 and EEGNet-8,2 models have similar performance – both the models converge to accuracies of 71%-72% after around 10 epochs, and their highest mean accuracies in the first 50 training epochs are 71.75% and 71.88%, respectively. The Sinc-ShallowNet model has better performance than the EEGNet models – it reaches a mean accuracy of 72.42% after 16 training epochs. The Conv-ShallowNet model has better performance than the Sinc-ShallowNet model. The model stabilizes on mean accuracies above 74% and its highest mean accuracy in the first 50 training epochs is 75.19%. The results indicate that the class-discriminative EEG features contained in this dataset could be well captured by shallow neural networks, since the both the InterpretableCNN and ShallowNet models have shallower structures compared to EEGNets.

We further investigate InterpretableCNN by comparing its performance with its variations. As it can be seen from the results shown in Fig. 2(b), the peak accuracy of InterpretableCNN is around 2% higher than that of 1DConv, where the separable convolution is replaced with standard 1-dimensional convolution. Other two variations NoPoinwise, where the pointwise convolutional layer is removed, and NoDepthwise, where the depthwise convolutional layer is removed, have quite similar performance as 1DConv. The obtained results indicate that the separable convolution can indeed boost the performance of the model, while the performance when using a standard 1D convolutional layer is similar to that when a single pointwise or a depthwise convolutional layer is used in the model. In addition, we can also observe that the performance of the model is largely affected when the batch normalization layer is removed. The obtained results indicate that robust performance of the proposed model is mainly the result of the compact design of the structure, while individual components including the separable convolution and batch normalization also have their positive contribution.

The accuracies of the conventional baseline methods with different classifiers are shown in Table II. We can see from the table that the mean classification accuracies obtained with different conventional methods range from 53.40%-72.68%. The highest mean accuracy of 72.68% is achieved by LogPower+GNB. The best accuracies for the baseline methods of RelativePower, PowerRatio and FourEntropies, are obtained with the SVM classifier, which are 68.64%, 64.24% and 66.49%, respectively. The best accuracy for the WaveletEntropy method is achieved with the LR classifier, which is 60.40%. We also notice that the band power related baseline methods of RelativePower, LogPower and PowerRatio have an overall better accuracy over the other two baseline methods using entropy features.

TABLE II. THE MEAN CROSS-SUBJECT CLASSIFICATION ACCURACIES (%) OF THE FIVE BASELINE METHODS COMBINED WITH DIFFERENT CLASSIFIERS.

| Classifiers | Relative-Power | Log-Power | Power-Ratio | Wavelet-Entropy | Four-Entropies |
|---|---|---|---|---|---|
| DT | 60.61 | 64.30 | 60.27 | 53.40 | 58.16 |
| RF | 64.76 | 69.54 | 63.39 | 56.69 | 61.82 |
| KNeighbors | 62.66 | 71.77 | 61.62 | 57.44 | 61.95 |
| GNB | 64.76 | **72.68** | 58.75 | 56.34 | 62.96 |
| LR | 68.58 | 70.24 | 63.17 | **60.40** | 60.62 |
| LDA | 66.29 | 70.44 | 64.19 | 59.71 | 60.98 |
| QDA | 65.00 | 61.62 | 59.19 | 59.37 | 57.40 |
| SVM | **68.64** | 71.95 | **64.24** | 60.18 | **66.49** |
| Mean | 65.16 | 69.07 | 61.85 | 57.94 | 61.30 |

### 2) Individual comparison results on the unbalanced dataset

In this section, we compare the best baseline methods with the proposed method on unbalanced data for each individual. Specifically, we conduct leave-one-subject-out cross validation, where the unbalanced EEG data from one subject is used for testing, while the balanced data from all other subjects are used for training in order to obtain the unbiased classifiers. Besides accuracy, we also use another two metrics, which are precision and recall, to evaluate the classification on the unbalanced data. The precision score is calculated by dividing the correctly classified drowsy samples by the total number of samples classified with the label of drowsiness. A low precision score indicates a high portion of alert samples are classified as drowsiness. The recall score is calculated by dividing the correctly classified drowsy samples by the total number of drowsy samples. A low recall score indicates a high portion of drowsy samples are classified as alertness.

We select two baseline deep learning models – the EEGNet-4,2 and Conv-ShallowNet, and two conventional baseline models–RelativePower+SVM and LogPower+GNB for comparison. We trained the deep learning models for 11 epochs after they have converged. As it can be seen from the results shown in Table III, the proposed model has the highest mean accuracy of 77.70% among the five models. By looking at individual classification results, we find that Subject 2 has a low accuracy for almost all the classifiers (except for the Conv-ShallowNet model). We further explore reasons behind the wrongly classified samples of the proposed model in the next section.

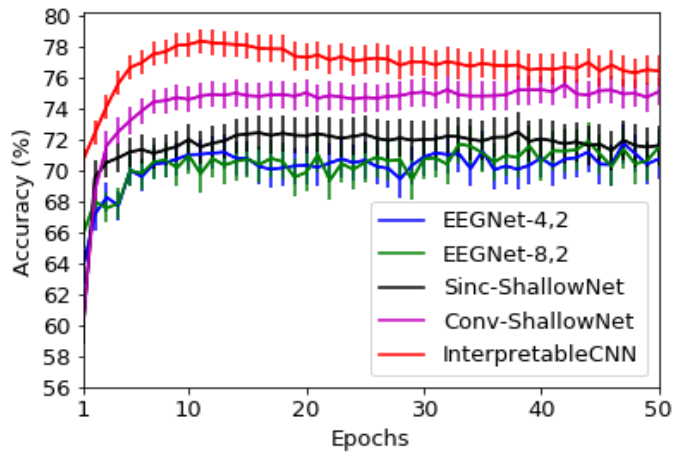

(a)

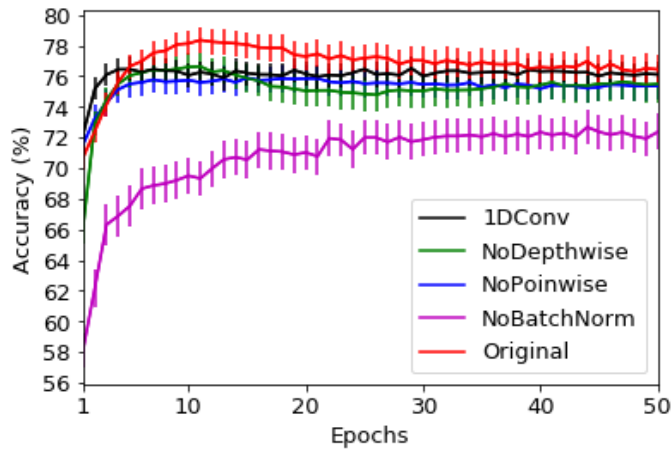

(b)

Fig. 2. Comparison of the mean cross-subject classification accuracies (%) between the proposed model InterpretableCNN and (a) benchmark deep learning models and (b) its variations, against training epochs from 1 to 50. The standard errors and accuracies averaged over 10 iterations for 11 subjects of each model are shown.

TABLE III. COMPARISON OF THE MEAN CROSS-SUBJECT ACCURACIES (%) ON THE UNBALANCED DATASET BETWEEN THE PROPOSED MODEL AND FOR BASELINE METHODS. THE PRECISION, RECALL, AND ACCURACIES OBTAINED FOR EACH SUBJECT ARE SHOWN IN THE TABLE.

| ID | EEGNet-4,2 | | | Conv-ShallowNet | | | RelativePower+SVM | | | LogPower+GNB | | | InterpretableCNN | | |
|---|---|---|---|---|---|---|---|---|---|---|---|---|---|---|---|
| | Pre. | Rec. | Acc. | Pre. | Rec. | Acc. | Pre. | Rec. | Acc. | Pre. | Rec. | Acc. | Pre. | Rec. | Acc. |
| 1 | 87.36 | 80.85 | 84.57 | 86.21 | 78.13 | 82.63 | 58.55 | 92.71 | 63.16 | 95.08 | 60.42 | 78.42 | 89.41 | 79.17 | **84.74** |
| 2 | 1.87 | 100.0 | 20.45 | 36.21 | 31.82 | **80.89** | 15.52 | 78.79 | 30.77 | 23.18 | 81.82 | 55.48 | 18.75 | 54.55 | 56.64 |
| 3 | 0.0 | 0.0 | 58.00 | 84.33 | 62.78 | 65.49 | 71.83 | 56.67 | 53.73 | 100.0 | 14.44 | 39.61 | 89.60 | 62.22 | **68.24** |
| 4 | 85.51 | 90.77 | **89.19** | 57.50 | 93.24 | 70.83 | 44.87 | 47.30 | 57.29 | 56.69 | 97.30 | 70.31 | 63.83 | 81.08 | 75.00 |
| 5 | 96.0 | 16.11 | 43.75 | 76.61 | 84.82 | 83.15 | 82.67 | 55.36 | 76.92 | 96.67 | 51.79 | 79.49 | 81.51 | 86.61 | **86.45** |
| 6 | 34.92 | 91.67 | 49.19 | 98.68 | 64.66 | 78.89 | 81.62 | 95.69 | **84.92** | 86.46 | 71.55 | 76.88 | 97.70 | 73.28 | 83.42 |
| 7 | 100.0 | 81.03 | **89.22** | 80.20 | 78.64 | 72.73 | 78.26 | 87.38 | 75.32 | 73.39 | 88.35 | 70.78 | 84.04 | 76.70 | 74.68 |
| 8 | 94.44 | 45.54 | 75.76 | 55.05 | 90.91 | 70.27 | 57.53 | 81.06 | 71.89 | 44.57 | 93.18 | 56.22 | 62.43 | 89.39 | **77.03** |
| 9 | 93.06 | 62.33 | 71.02 | 89.03 | 87.90 | **91.0** | 81.25 | 74.52 | 83.25 | 83.44 | 86.62 | 88.00 | 86.93 | 84.71 | 89.00 |
| 10 | 26.67 | 100.0 | 89.81 | 100.0 | 22.22 | 82.93 | 85.37 | 64.81 | **89.84** | 72.97 | 50.00 | 84.96 | 70.37 | 70.37 | 86.99 |
| 11 | 69.75 | 90.22 | **80.09** | 70.18 | 61.07 | 65.16 | 60.80 | 92.37 | 63.93 | 66.67 | 73.28 | 65.98 | 76.67 | 70.23 | 72.54 |
| Ave. | 62.69 | 68.96 | 68.28 | 75.82 | 68.74 | 76.72 | 65.30 | 75.15 | 68.28 | 72.65 | 69.89 | 69.65 | 74.66 | 75.30 | **77.70** |

## B. Interpretation on the learned characteristics from EEG signals

Deriving insights into what the model has learned from the data is an important procedure of model validation. In this section, we investigate what patterns have been learned by the model to distinguish between alert and drowsy EEG signals with the interpretation method proposed in Section III.C. In addition, we also explore the reasons behind some wrongly classified samples for selected subjects discussed in the previous section.

We display some representative samples from different subjects that are correctly classified with high likelihoods of drowsy and alert signals in Fig. 3 and Fig. 4, respectively. We have found that most EEG samples classified as drowsy signals by the model with a high confidence level commonly contain a high portion of Theta waves, e.g., Fig. 3(a), or Alpha waves, e.g., Fig. 3(b-d). For the first sample shown in Fig. 3(a), it can be observed that the model has identified several episodes that contain rhythmic bursts in the Theta band as strong evidence of the drowsy state. These bursts in the Theta band, or called "drowsy bursts", have been found to frequently appear in EEG signals during drowsiness [56]. For the samples shown in Fig. 3(b-d), it can be observed that the model has identified spindle-like structures in Alpha frequency from several episodes of the signal as indicators of drowsiness. Actually, the captured Alpha spindles, which can be characterized by a narrow frequency peak within the alpha band [57], have been found to be strong indicators of early drowsiness and used in various driving simulator studies to identify the driver drowsiness [58]. We have also observed another interesting pattern that the central and centro-parietal EEG channels, e.g., Cz in Fig. 3(b), usually play a more importance role than peripheral channels for drowsiness classification. One possible reason is that these channels mostly contain cleaner cortical signals, where the drowsiness-related features are more distinguishable than those from the peripheral or frontal channels, which are more likely to be contaminated by artifacts caused by brain muscle tension [59] or eye movements.

As it can be seen in Fig. 4, we have found that the samples classified with a high likelihood of the alert signals commonly contain more artifacts than samples shown in Fig. 3. For the first two samples shown in Fig. 4(a) and 4(b), the model has identified several episodes of the signals from peripheral EEG channels (including F7, FT7, T5, O1, O2 and TP8), which result in high relative power in the Beta frequency band, as evidence for the classification. Actually, the identified Beta waves are mixtures of local cortical Beta waves, which are often associated with active, busy or anxious thinking or active concentration [60] and electromyography (EMG) activities, which have peaks in the Beta frequency range that resemble EEG Beta peaks causing the greatest contamination on EEG signals at the periphery of the scalp near the active muscles [59]. For the samples shown in Fig. 4(c) and 4(d), it can be observed that the model has identified several episodes that contain large voltage change of signals from frontal EEG channels as evidence of alertness. These large-amplitude and low-frequency waves, resulting a high power in the Delta frequency band, are caused by eye blinks and eye movement activities during the wakeful state. Actually, it is out of our expectation that the model uses features that are commonly regarded as artifacts contained in EEG as indicators of the alert state. In fact, these artifacts including EMG and eye movement activities are the strongest components of wakeful EEG signals [56], so that it makes sense to some extent that the model has identified such features from group statistics of the data.

Following the discussion in Section IV.A.2), we investigate the reasons behind the wrongly classified samples from individual subjects with the proposed visualization technique. For Subject 2, we find that sensor noise contained in the EEG signals is one of the major reasons that lead to the low classification accuracy. As it can be seen in the example shown in Fig. 5(a), the model has falsely identified the noise, which causes significant fluctuations in the TP7 channel, as evidence of the alert state. Despite most of such sensor noise have been filtered out in the pre-processing phase, there are still some left that negatively affect the classification. In addition, we have also noticed that many alert samples from Subject 2 resemble typical drowsy EEG signals, e.g., samples from Fig. 3. As it can be seen from Fig. 5(b), the sample does not contain any apparent alertness-related features. On the contrary, it contains spindle-like structures in almost all channels, which cause a high relative power in the Alpha frequency band from the occipital channels. The model has justified its classification by localizing several episodes from the sample that contain such drowsiness-related spindles. One possible explanation is that the subject was already in the early drowsy stage but still behaving normally when these samples were captured.

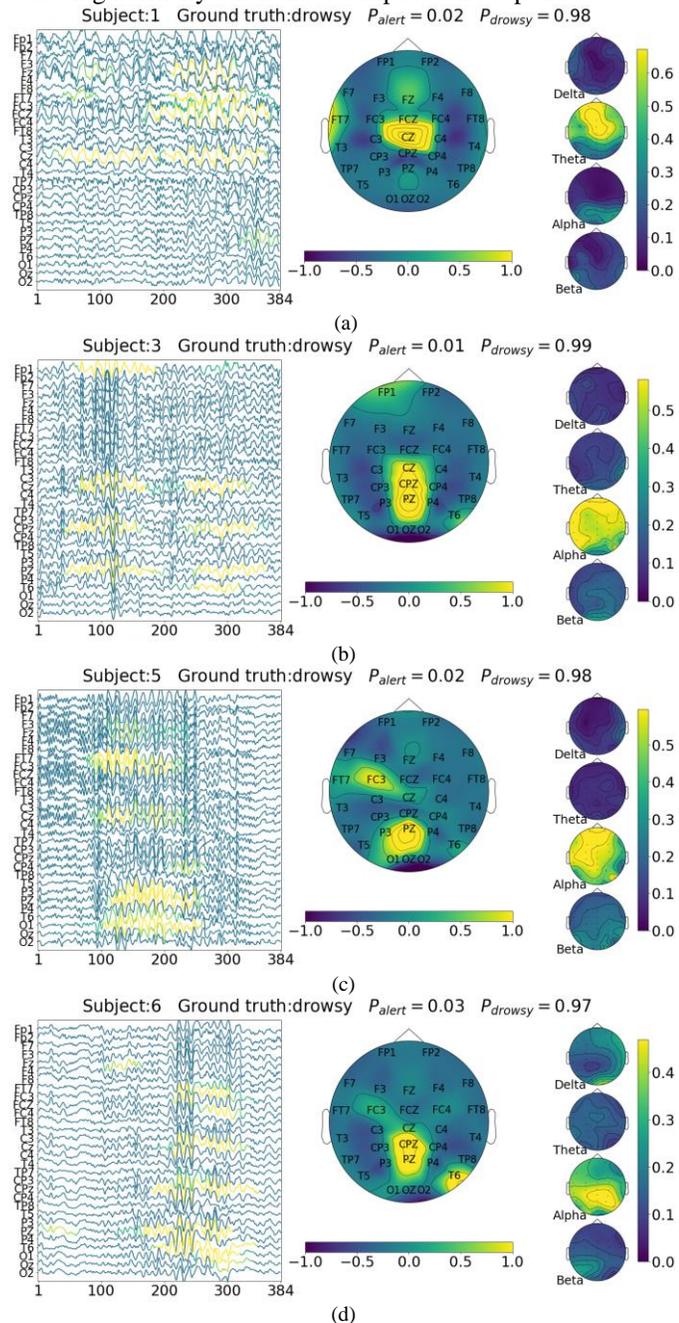

Fig. 3. Visualization of learned patterns on selected drowsy EEG samples that are correctly classified by the network with high likelihood. The subject ID, ground truth label, likelihood output by the model for alert and drowsy classes are shown on top of each sub-figures. In the left part of each sub-figure, the contributing regions to classification are highlighted by the heatmap overlaid on the input EEG signal, which is obtained with the proposed interpretation technique described in Section III.C. The topologic heatmap in the middle of each sub-figure is obtained by averaging the heatmap over each EEG channel. It summarizes to which extent each channel contributes to the final classification. The relative powers of Delta, Theta, Alpha and Beta frequency bands for each EEG channel of the input signal is shown in the right part of each sub-figure.

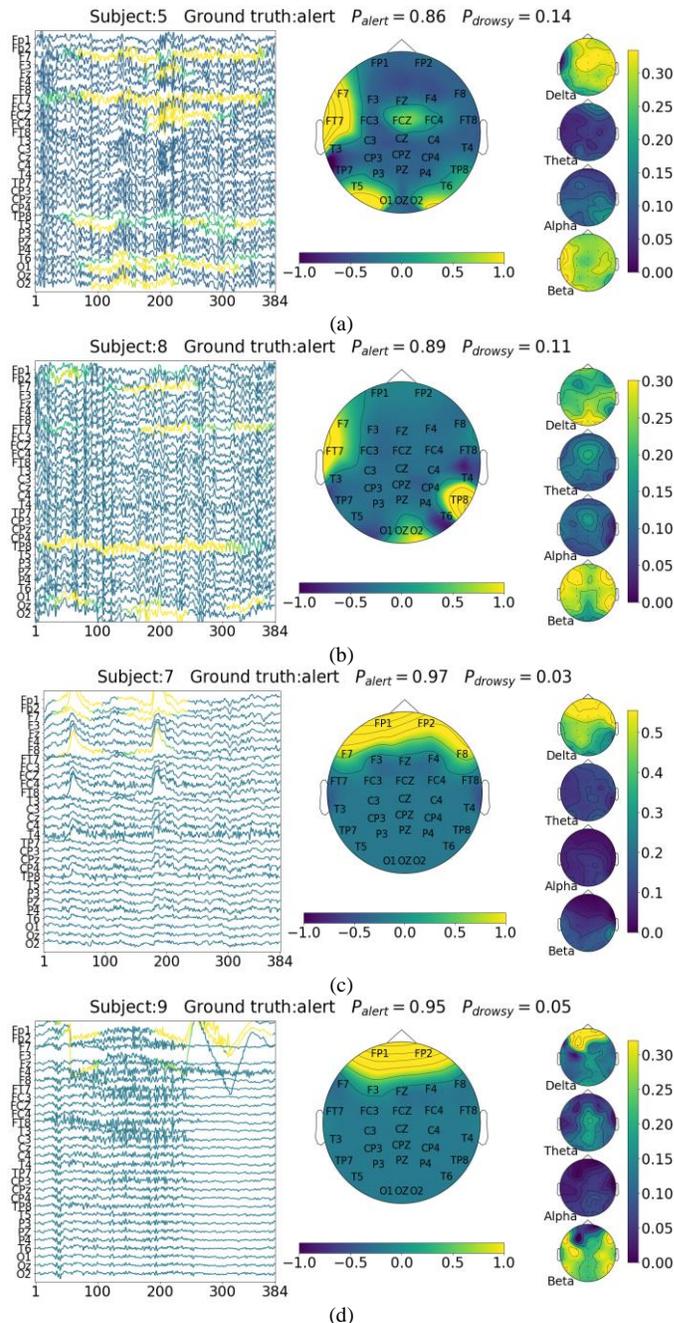

Fig. 4. Visualization of learned patterns on selected alert EEG samples that are correctly classified by the network with high confidence.

We have also found that the EMG activities could be a common factor that affects the classification across different subjects. Although the EMG activities are not very common in typical drowsy EEG signals, as it can be seen from samples shown in Fig. 3, their occasional appearance in drowsy EEG signals could be quite misleading and affect the decision of the model. From the three representative samples shown in Fig. 6, we can observe that the model has falsely identified several episodes containing EMG activities from peripheral EEG channels, e.g., F7, FT7, as evidence for its classification, regardless of the apparent drowsiness-related Alpha spindles contained in the samples.

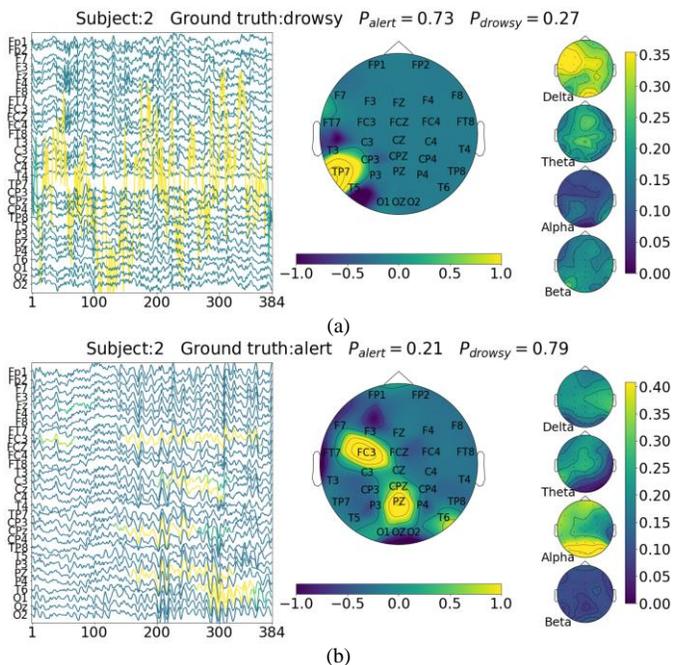

Fig. 5. Visualization of learned patterns on selected wrongly classified samples from Subject 2.

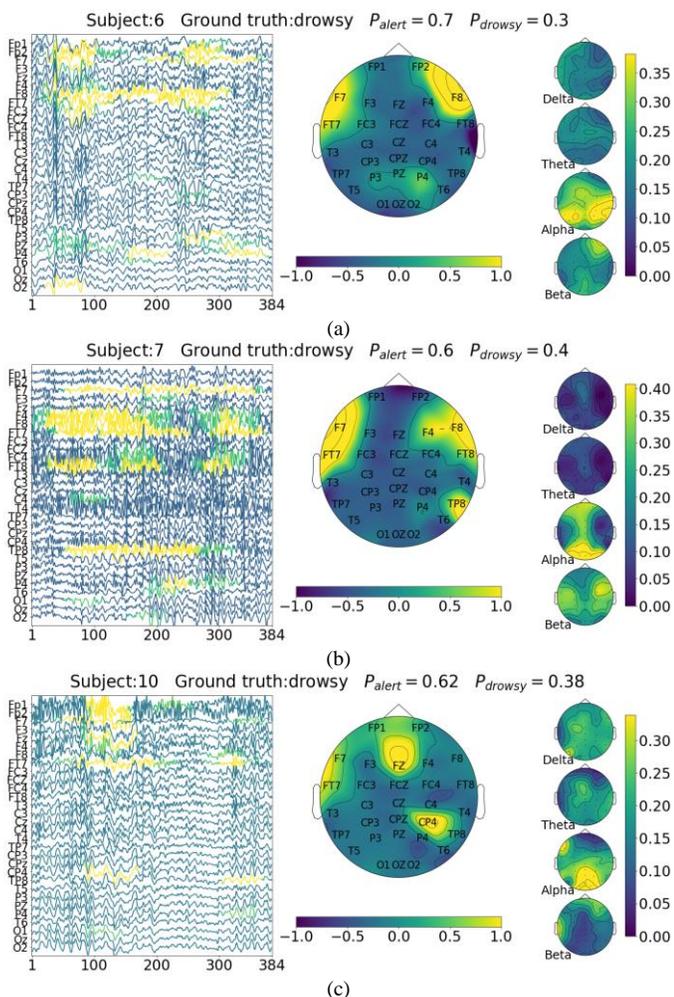

Fig. 6. Visualization of learned patterns on selected wrongly classified samples from three different subjects.

## V. DISCUSSION AND FUTURE WORKS

In this paper, we consider a promising topic of using interpretable deep learning models to discover meaningful patterns related to different mental states from cross-subject EEG signals. The model has a compact structure and it uses separable convolutions to process the EEG signals in a spatial-temporal sequence. In order to allow the model to "explain" its decisions, we designed an interpretation technique, which is inspired by the CAM and Fixation-CNN methods, to reveal the important local regions of the sample for the classification.

By sample-wise analysis of the classification results, we found the model had learned biologically meaningful features from the EEG signals to distinguish between alert and drowsy states. Specifically, the model has discovered neurologically interpretable features, such as Alpha spindles and bursts in Theta band, as evidence of drowsiness. It has also learned to recognize eye blink and eye movement features, as well as the Beta waves from peripheral EEG channels, which are mixtures of cortical signals and EMG activities, as evidence of alertness. The proposed method has advantages over conventional methods by revealing how different components in EEG signals are related to different mental states directly from the raw high-dimensional EEG signals. For example, to our knowledge this is the first study showing the EMG activities from peripheral EEG channels, which are commonly regarded as artifacts to be removed in the pre-processing phase, could be beneficial to distinguish between alert and drowsy EEG signals.

In addition, we also explored the reasons behind some wrongly classified samples in order to understand how the cross-subject classification accuracy can be further improved. The obtained results indicate it is necessary to carefully design the pre-processing pipeline for dealing with different kinds of artifacts and noise in the signals according to their impacts on the classification task. For our scenario, the sensor noise that causes significant fluctuations of the signals should be eliminated since it negatively affects the classification, while the eye movements/blinks and EMG activities could be discriminatively treated considering they have been learned by the model as typical indicators of alertness. In addition, the labeling of samples could be further improved, since the performance or behaviors of subjects, e.g., reaction time, may not faithfully reflect the actual mental states of subjects in certain circumstances. Instead of merely using thresholds hard-coded on behavior/performance metrics of the subjects, it could be a promising way of incorporating the "model explanation" into the labeling process to reduce bias of the sample labeling. With regard to the cases where the EMG activities mislead that model by diverting its attention from important spindle-like features for making the decision, e.g., samples from Fig 6, a potential solution is augmenting the training data of this kind with artificial samples created by generative models, in order for the model to learn the priority of these contradictory clues when they are simultaneously present in a sample.

We have to admit that the EEG signals are in nature not as easy to interpret as other forms of data, e.g., images and natural languages. Only a small portion of the samples were interpreted in this paper as a pilot study. More efforts in future work will be made to interpret the deep learning models and design methods accordingly to solve the faced problems towards getting calibration-free brain computer interfaces.

## VI. CONCLUSION

In this paper, we developed a novel CNN model for the purpose of discovering common patterns related to different mental states in EEG signals across different subjects. The model has a compact structure and it uses separable convolutions to process the EEG signals in a spatial-temporal sequence. In addition, we also designed interpretation techniques to reveal what has been learned by the model for classification by highlighting the relevant parts of the input signal. Results show that the proposed model has better performance than both conventional baseline methods and state-of-the-art deep learning models for cross-subject drowsiness recognition. The interpretation results show that the model has learned to identify biologically meaningful features, e.g., Alpha spindles, from the data and use them as evidence to distinguish between drowsy and alert EEG signals. In addition, we also explored reasons behind some wrongly classified samples and discussed how these problems could be potentially solved. Our work illustrates a promising direction to discover meaningful patterns related to different mental states from complex EEG signals towards calibration-free brain computer interfaces.

### ACKNOWLEDGMENT

This research is supported by the National Research Foundation, Singapore under its International Research Centres in Singapore Funding Initiative. Any opinions, findings and conclusions or recommendations expressed in this material are those of the author(s) and do not reflect the views of National Research Foundation, Singapore.